# Baseline drift effect on the performance of neutron and γ ray discrimination using frequency gradient analysis

LIU Guo-Fu(刘国福)　　LUO Xiao-Liang(罗晓亮)　　Yang Jun(杨俊)　　LIN Cun-Bao(林存宝)
Hu Qing-Qing (胡青青)　　Peng Jin-Xian(彭进先)

Department of Instrument Science and Technology, National University of Defense Technology, Changsha 410073, China

**Abstract:** Frequency gradient analysis (FGA) effectively discriminates neutrons and γ rays by examining the frequency-domain features of the photomultiplier tube anode signal. This approach is insensitive to noise but is inevitably affected by the baseline drift, similar to other pulse shape discrimination methods. The baseline drift effect is attributed to **the** factors such as power line fluctuation, dark current, noise disturbances, hum, and pulse tail in front-end electronics. This effect needs to be elucidated and quantified before the baseline shift can be estimated and removed from the captured signal. Therefore, the effect of baseline shift on the discrimination performance of neutrons and γ rays with organic scintillation detectors using FGA is investigated in this paper. The relationship between the baseline shift and discrimination parameters of FGA is derived and verified by an experimental system consisting of an americium–beryllium source, a BC501A liquid scintillator detector, and a 5 GSample/s 8-bit oscilloscope. Both theoretical and experimental results show that the estimation of the baseline shift is necessary, and the removal of baseline drift from the pulse shapes can improve the discrimination performance of FGA.

## 1. Introduction

Given the distinct differences between the frequency spectrum of the γ-ray and neutron signals, which can be used as prominent features to discriminate them, a novel n/γ discrimination method called the frequency gradient analysis (FGA) method has been proposed and implemented [1,2]. In Ref. [1], the performance of the FGA method and the pulse gradient analysis (PGA) method [3] has been studied and compared on a theoretical basis and then verified by the time-of flight (TOF) method. In Ref. [2], a comparison of discrimination performance of the FGA method and the conventional charge comparison (CC) method has been studied and verified by TOF.

The results from both papers show that the FGA method exhibit a strong insensitivity to the variation in pulse response of the photomultiplier tube (PMT), which can be used to discriminate neutron and γ-ray events in a mixed radiation field, and that the FGA method has the potential to be implemented in current embedded electronics systems to provide real-time discrimination in standalone instruments.

However, this approach is inevitably affected, like other pulse shape discrimination (PSD) methods [3-9], by the baseline drift due to factors such as power line fluctuation, dark current, noise disturbances, hum and pulse tail in front-end electronics [10,11]. Although we all know that it is necessary to cancel the baseline drift to improve the discrimination performance and a number of baseline restoration (BLR) approaches, from classic analog methods to digital filter solutions, have been proposed to reduce the effect of the baseline drift [12], the relationship between the baseline shift and the discrimination parameters extracted from the pulse shape has not yet been quantitatively addressed.

To further improve the discrimination performance of FGA, decrease the complexity of the BLR approaches and reduce the cost on calculation, the effect of the baseline drift on FGA has been thoroughly investigated by theoretical analysis and experimental verification in this study. A detailed description of the experimental environment is provided in Section 2. The relationship between the baseline shift and discrimination parameter extracted from the pulse shape is quantitatively derived in Section 3. The experimental results are provided and discussed in Section 4, followed by the conclusions derived from this research.

## 2. The experimental method

The experimental data analyzed in this work were acquired using a radiation measurement system at the Institute of Nuclear Physics and Chemistry, the Chinese Academy of Engineering Physics, Mianyang, China. The schematic of the experimental setup is shown in Fig. 1. A BC501A organic liquid scintillator was exposed to a mixed radiation field produced by a $^{241}$Am-Be neutron source suspended on a three-legged stand. The neutron source was positioned 1750 mm away from the ground and with a distance of 1150 mm to the scintillation detector at the same height. The liquid scintillation detector consisted of a Φ50.8 mm × 50.8 mm cylindrical cell scintillation detector filled with BC501A organic liquid, optically coupled to an EMI 9807B PMT operated with a negative supply voltage of -1400 V DC. The output signal from the liquid scintillator was transmitted to Channel 1 of a Tektronix digital phosphor oscilloscope (2.5 GSample/s and 8-bit resolution) by approximately 25 m of the high bandwidth cable. The data were then streamed to a PC. Each pulse shape consisted of 1000 samples taken at 0.4 ns intervals, and approximately 10000 digitized events were collected.

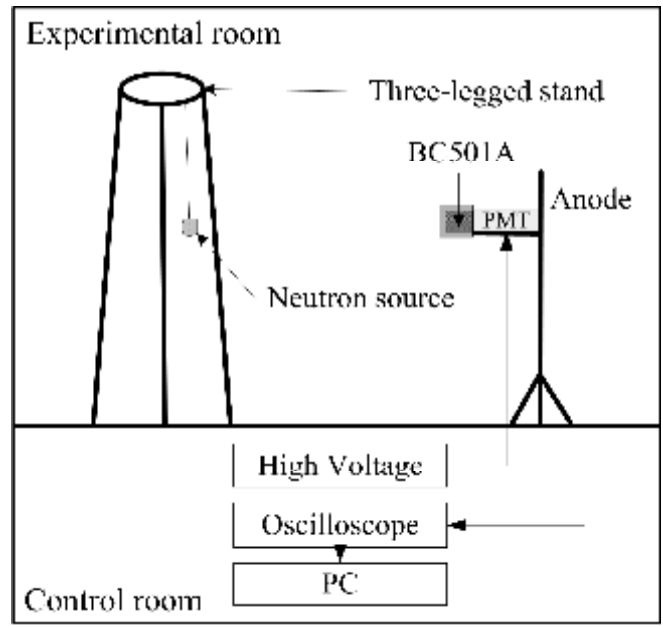

**Fig. 1** Schematic of the experimental setup consisting of the $^{241}$Am-Be neutron source, liquid scintillator detector with a BC501A, and a 5 GSample/s 8-bit oscilloscope. The location of the source, detector, and cables are not to scale.

## 3. Analysis of the baseline drift effect on FGA

### 3.1. Realization of FGA

Considering that the principle of FGA has been given in detail in Ref. [1], only the realization of this method is described in the current study. We also provide a schematic of the data process of a typical event obtained in the experiment in Fig. 2. The procedure is described below.

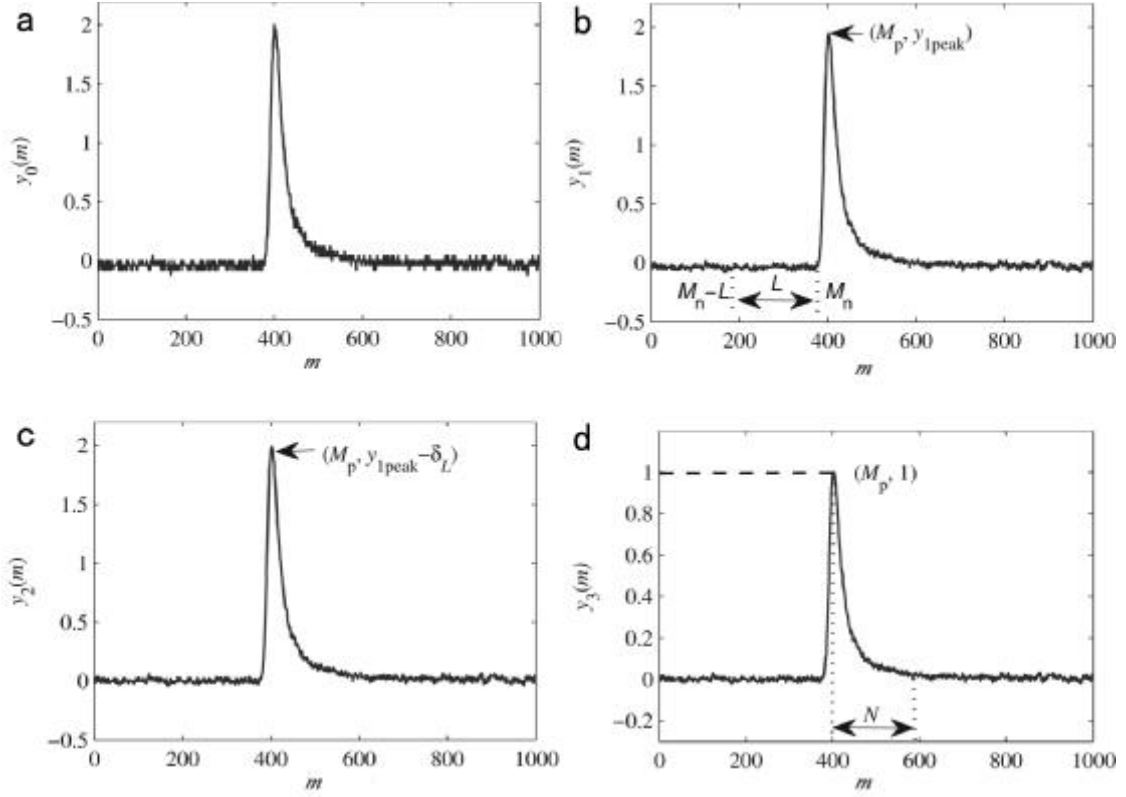

**Fig. 2** Demonstration of data pre-processing of a typical event (either γ-ray or neutron event): (a) the original pulse $y_0(m)$, (b) the filtered pulse $y_1(m)$, (c) the pulse $y_2(m)$ with baseline shift being subtracted, and (d) the normalized pulse. In the figure, the integer variable *m* is the discrete time index.

(1) The captured original waveform is $y_0(m)$ (Fig. 2a), with a sampling frequency denoted by $F_s$, and the data length is $M$. To reduce the effect of noise, we first process $y_0(m)$ using a moving-average (MA) filter, i.e.

$$y_1(m) = \sum_{k=0}^{q} b_k y_0(m-k), \tag{1}$$

where *m* and *k* are the discrete-time indexes and $b_k$ ($k=0,1,\ldots q$) is the coefficient of the MA filter. In practice, we usually let $q=1$ and $b_0 = b_1 = 0.5$ to reduce the processing cost. Eq. (1) then becomes

$$y_1(m) = \frac{1}{2}[y_0(m) + y_0(m-1)]. \tag{2}$$

(2) A programmable length mean filter is used to estimate the baseline shift *d* (Fig. 2b), i.e.

$$d_L = \frac{1}{L} \sum_{k=M_n-L+1}^{M_n} y_1(k), \tag{3}$$

where $M_n$ is the last data point of the baseline data and determined by the trigger level and the pulse's rise-time, after which the captured sample is considered as the pulse data. $L$ is the length of the truncated baseline data and $d_L$ is an estimation of $d$.

(3) After $d_L$ is determined, we can subtract it from the preprocessed waveform (Fig. 2c), i.e.

$$y_2(m) = y_1(m) - d_L. \tag{4}$$

Assuming that $y_{1peak} = \max[y_1]$, then $\max[y_2(m)] = \max[y_1(m) - d_L] = y_{1peak} - d_L$.

(4) The new waveform $y_2$ is normalized to obtain

$$y_3(m) = \frac{y_2(m)}{\max[y_2(m)]} = \frac{y_1(m) - d_L}{y_{1peak} - d_L}. \tag{5}$$

(5) The normalized waveform is truncated to obtain the falling portion of the pulse. Assuming that the data length of this part is $N$ (Fig. 2d), we obtain

$$x(n) = y_3(n + M_p) = \frac{y_1(n + M_p) - d_L}{y_{1peak} - d_L}, \tag{6}$$

where $M_p$ is the data point at which $y_3$ is the maximum.

(6) The zero-frequency component $X(0)$ and the first frequency component $X(1)$ of the Fourier series of $x(n)$ are calculated according to the following equations [13]:

$$X(0) = \sum_{n=0}^{N-1} x(n) = N\overline{x}, \tag{7a}$$

$$X(1) = \sum_{n=0}^{N-1} x(n)\cos(\frac{2\pi}{N}n) - i \cdot \sum_{n=0}^{N-1} x(n)\sin(\frac{2\pi}{N}n), \tag{7b}$$

where $\overline{x}$ is the mean value of $x(n)$. If the values of $\cos(2\pi n/N)$ and $\sin(2\pi n/N)$ are calculated in advance, $X(1)$ can be obtained quickly using a lookup table.

(7) The discrimination parameter $k$ is calculated as
$$k = |X(0)| - |X(1)|. \tag{8}$$

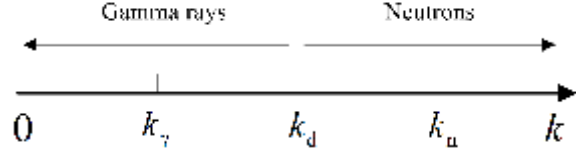

**Fig. 3** Schematic of the differentiation of the neutron and γ-ray event based on the $k$ value.

According to the discrimination principle of FGA [1], the discrimination parameter for γ-ray events is smaller than that for neutron events. Therefore, we can differentiate neutron events from γ-ray events by the $k$ value, which is calculated according to Eq. (8). The schematic of this process is shown in Fig. 3, where $k_\gamma$ and $k_n$ are the ideal frequency gradients of γ rays and neutrons, respectively, and $k_d$ is a discrimination value that is determined from the experimental data. In fact, the $k$ value for each waveform has an average value close to $k_\gamma$ and $k_n$ for γ rays and neutrons, respectively. However, a spread of the $k$ values was found because of a number of potential sources of fluctuation in the detector response. Comparing $k$ with $k_d$, we can classify the processed event as a γ-ray or a neutron event.

### 3.2. Relationship between the baseline shift and discrimination parameter

To study the baseline shift effect on the discrimination performance, we substitute Eq. (6) into Eqs. (7a) and (7b), to obtain

$$X(0) = \frac{1}{y_{1peak} - d_L} \sum_{n=0}^{N-1} y_1(n+M_p) - \frac{Nd_L}{y_{1peak} - d_L}, \tag{9a}$$

$$X(1) = \frac{1}{y_{1peak} - d_L}[\sum_{n=0}^{N-1} y_1(n+M_p)\cos(\frac{2\pi}{N}n) - i \cdot \sum_{n=0}^{N-1} y_1(n+M_p)\sin(\frac{2\pi}{N}n)]. \tag{9b}$$

Here, when deducing $X(1)$, we use the constant equation $\sum_{n=0}^{N-1}\exp(-i\frac{2\pi}{N}n) = 0$.

With $l = d_L / y_{1peak}$, Eqs. (9a) and (9b) becomes

$$X(0) = \frac{1}{(1-l)}\sum_{n=0}^{N-1} \frac{y_1(n+M_p)}{y_{1peak}} - N\frac{l}{1-l}, \tag{10a}$$

$$X(1) = \frac{1}{(1-l)}[\sum_{n=0}^{N-1} \frac{y_1(n+M_p)}{y_{1peak}} \cos(\frac{2\pi}{N}n) - i \cdot \sum_{n=0}^{N-1} \frac{y_1(n+M_p)}{y_{1peak}} \sin(\frac{2\pi}{N}n)]. \qquad (10b)$$

Substituting Eqs. (10a) and (10b) into Eq. (8), we can obtain the relationship between $l$ and $k$

$$k_l = |X(0)| - |X(1)| = [\frac{1}{(1-l)} \sum_{n=0}^{N-1} \frac{y_1(n+M_p)}{y_{1peak}} - N\frac{l}{1-l}]$$
$$- \frac{1}{(1-l)} |\sum_{n=0}^{N-1} \frac{y_1(n+M_p)}{y_{1peak}} \cos(\frac{2\pi}{N}n) - i \cdot \sum_{n=0}^{N-1} \frac{y_1(n+M_p)}{y_{1peak}} \sin(\frac{2\pi}{N}n)|. \qquad (11)$$

### 3.3. Error analysis of $k_l$

Supposing that the true baseline shift is $d_t$ and $l_t = d_t / y_{1peak}$, then the true discrimination parameter $k_t$ can be obtained from Eq. (11) by substituting $l$ with $l_t$. Hence, the difference between $k_l$ and $k_t$ is

$$\Delta k_l = k_l - k_t = [\frac{1}{(1-l)} - \frac{1}{(1-l_t)}] \sum_{n=0}^{N-1} \frac{y_1(n+M_p)}{y_{1peak}} - N[\frac{l}{(1-l)} - \frac{l_t}{(1-l_t)}]$$
$$- [\frac{1}{(1-l)} - \frac{1}{(1-l_t)}] |\sum_{n=0}^{N-1} \frac{y_1(n+M_p)}{y_{1peak}} \cos(\frac{2\pi}{N}n) - i \cdot \sum_{n=0}^{N-1} \frac{y_1(n+M_p)}{y_{1peak}} \sin(\frac{2\pi}{N}n)|. \qquad (12)$$

The values of $l$ and $l_t$ are usually much smaller than one, which results in $\frac{1}{1-l} - \frac{1}{1-l_t} \approx 0$, and the values of the first and third terms on the right side of Eq. (12) are both positive. Thus, the difference between the two values is also almost equal to zero and we can neglect these two items. Thus, the expression for the difference between the calculated and true $k$ values becomes

$$\Delta k_l = k_l - k_t \approx -N[\frac{l}{(1-l)} - \frac{l_t}{(1-l_t)}] \approx N(l_t - l) = \frac{N}{y_{1peak}}(d_t - d_L). \qquad (13)$$

From Eq. (13), a more accurate estimation of the baseline shift $d_L$ results in a smaller error between $k_l$ and $k_t$, which leads to a better discrimination performance of neutrons and γ rays. According to Eq. (3), $d_L$ is mainly determined by the truncated noise data length $L$.

## 4. Results and discussion

In this section, we first calculate the discrimination parameter of the pulse shape shown in Fig. 2 for different values of the baseline shift $d_L$, by changing the truncated noise data length $L$ in Eq. (3). The discrimination results of all events are then given. Finally, we discuss the effect of the baseline shift on the discrimination performance of FGA. In the implementation of the FGA algorithm, we select the data length of the falling portion of the pulse to be 256, i.e. $N = 256$.

According to the total yields of neutrons and γ rays of the Am-Be neutron source, the geometric structure and relative position of the neutron source and the detector, the estimated count rate is approximately 127/s. The probability for the pile-up of the signals is very small. Therefore, the effect of the pile-up on the baseline becomes negligible in the subsequent calculations and discussions.

### 4.1. Calculation of discrimination parameter of a single event

To demonstrate clearly the effect of the baseline shift on the discrimination parameter of FGA, we consider the pulse shown in Fig. 2 as an example. The calculated results are shown in Table 1 for different values of $L$. We selected 128 as the largest $L$ value because we have also tested some larger values of $L$, such as $L = 160$, $L = 192$, or $L = 256$, but found that the figure-of-merit (*FOM*) values were almost similar. Hence, we used $L = 128$ for simplicity. The other advantage of choosing $L = 128$ is that the effect of pile-up events can be reduced.

In Table 1, $L = 0$ indicates that we do not process the baseline shift, and therefore $d_L = 0$ in Eq. (4). If we assume that the distribution of the baseline shift is approximately a Gaussian white noise with a mean value of $d_t$, larger values of $L$ give a small difference between $d_L$ and $d_t$. Therefore, we set $d_t$ and $k_t$ to the values obtained for the largest $L$ used in this work, i.e. $L = 128$, $d_t = d_L |_{L=128}$, and $k_t = k_I |_{L=128}$. The results shown in Table 1 are consistent with the analyses given in Section 3.

**Table 1** The calculated parameters with the event shown in Fig. 2 under different data length $L$.

| $L$ | $d_L$ | $l$ | $k_I$ | $\Delta k_I$ |
|---|---|---|---|---|
| 0 | 0 | 0 | 6.68±0.01 | -4.892±0.200 |
| 8 | -0.0446±0.0064 | 0.0229±0.0032 | 12.3±0.8 | 0.696±0.800 |
| 16 | -0.0447±0.0038 | -0.0230±0.0020 | 12.2±0.5 | 0.714±0.500 |
| 32 | -0.0417±0.0027 | -0.0214±0.0013 | 11.9±0.4 | 0.339±0.400 |
| 64 | -0.0410±0.0018 | -0.0211±0.0009 | 11.8±0.2 | 0.257±0.300 |
| 128 | -0.0389±0.0013 | -0.0200±0.0006 | 11.6±0.2 | 0 |

### 4.2. Discrimination result under different baseline shift

With the FGA algorithm described in Section 3.1, the scatter plots of peak amplitude against discrimination parameter for different baseline shifts are shown in Fig. 4. A total of 5000 events are used for this analysis. As explained in Section 3.1, given the slower decay rate of a neutron-induced pulse, it has a higher discrimination value than a γ-ray pulse for the same peak amplitude. Hence, the neutron and γ-ray distributions are separated in terms of the discrimination parameter.

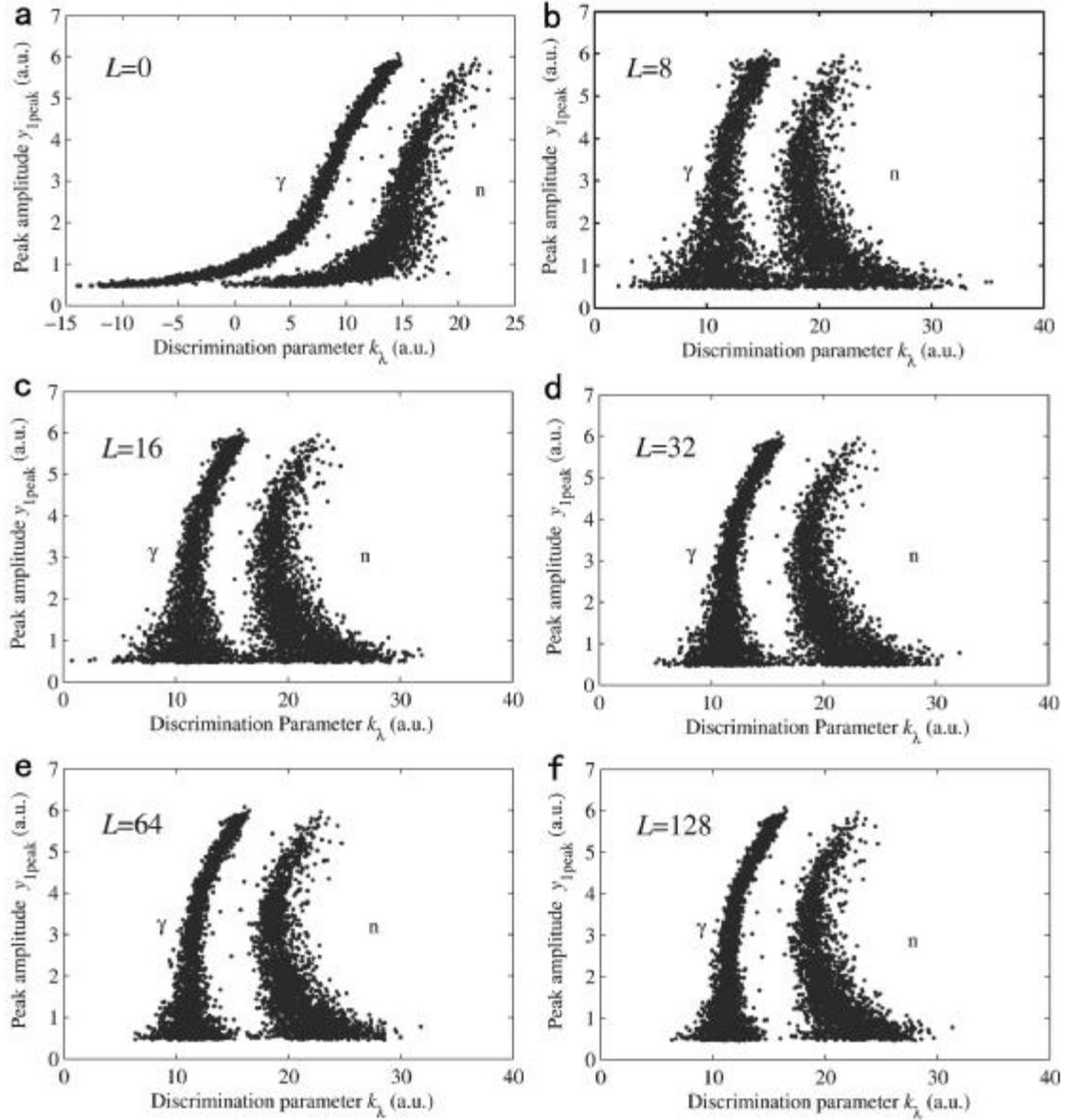

**Fig. 4** Scatter plots of peak amplitude versus discrimination parameter for different baseline shifts calculated by Eq. (3) and different values of $L$. The trigger threshold value is approximately 0.5, corresponding to the light output of approximately 350 keV of electron-equivalent recoil energy (keVee).

### 4.3. Comparison of discrimination performance

To evaluate the separation of the neutron and γ-ray distributions and compare the discrimination performance under different baseline shifts, the corresponding fitted Gaussian distributions of the FGA data under different baseline shifts are shown in Fig. 5. The *FOM* values for each baseline shift are calculated using Eq. (14):

$$FOM = \frac{S}{FWHM_\gamma + FWHM_n}, \qquad (14)$$

where $S$ is the separation between the centroids of the neutron and the γ-ray peaks in the $k_l$ spectrum. $FWHM_\gamma$ and $FWHM_n$ are the full-width-at-half-maximum values of the γ-ray and neutron peaks, respectively [14]. If the probability distribution function of each event is consistent with a Gaussian distribution, Eq. (14) becomes

$$FOM = \frac{|m_n - m_\gamma|}{2.35(s_n + s_\gamma)}, \quad (15)$$

where $m_\gamma$ and $m_n$ are the means of the γ-ray and neutron Gaussians, respectively. The standard deviation, $s$, is given as $s_\gamma$ and $s_n$ for the γ-ray and neutron Gaussians, respectively. The values of these parameters and the corresponding $FOM$ values under different baseline shifts can be calculated from the experimental results of FGA using the method given in Ref. [1], and the results are listed in Table 2.

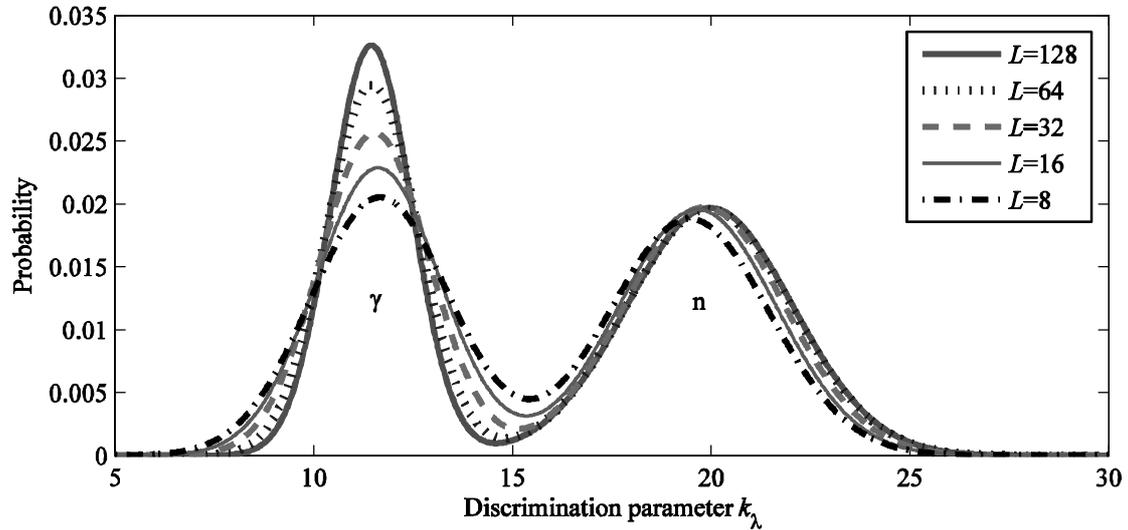

**Fig. 5** Corresponding fitted Gaussian distributions for the FGA data shown in Fig. 4 under different baseline shifts.

**Table 2** Means and errors of the γ ray and neutron Gaussians calculated from the experimental results using FGA and the corresponding *FOM* values under different baseline shifts.

| $L$ | $m_\gamma$ | $s_\gamma$ | $m_n$ | $s_n$ | *FOM* |
|---|---|---|---|---|---|
| 8 | 11.7±0.1 | 1.74±0.10 | 19.4±0.1 | 2.00±0.10 | 0.884±0.040 |
| 16 | 11.7±0.1 | 1.56±0.10 | 19.6±0.1 | 1.98±0.11 | 0.965±0.043 |
| 32 | 11.5±0.1 | 1.37±0.08 | 19.8±0.1 | 2.01±0.12 | 1.04±0.05 |
| 64 | 11.5±0.1 | 1.15±0.07 | 19.9+0.1 | 2.07±0.12 | 1.12±0.05 |
| 128 | 11.5±0.1 | 1.02±0.06 | 19.9±0.1 | 2.08±0.12 | 1.17±0.05 |

The effect of the baseline shift on the discrimination performance is qualitatively shown in Figs. 4 and 5 and quantitatively in Table 2. The results demonstrate that with a larger $L$, the distance between the neutron and the γ-ray branches becomes longer. Then, the overlapped region of these two branches becomes smaller, and the *FOM* becomes larger, indicating the enhanced discrimination performance. These results are consistent with the conclusions derived from Eq. (13). With a larger $L$ value, the error between the estimated $d_L$ and the true value $d_t$ becomes smaller, resulting in a smaller $\Delta k_1$, accordingly.

According to Eq. (13), $\Delta k_1$ is not only proportional to $(d_t - d_L)$, i.e. the error between the estimated $d_L$ and the true value $d_t$, but also inversely proportional to $y_{1peak}$, i.e. the peak amplitude of the captured signal. A smaller peak amplitude results in a larger $\Delta k_1$, which leads to decreased discrimination performance.

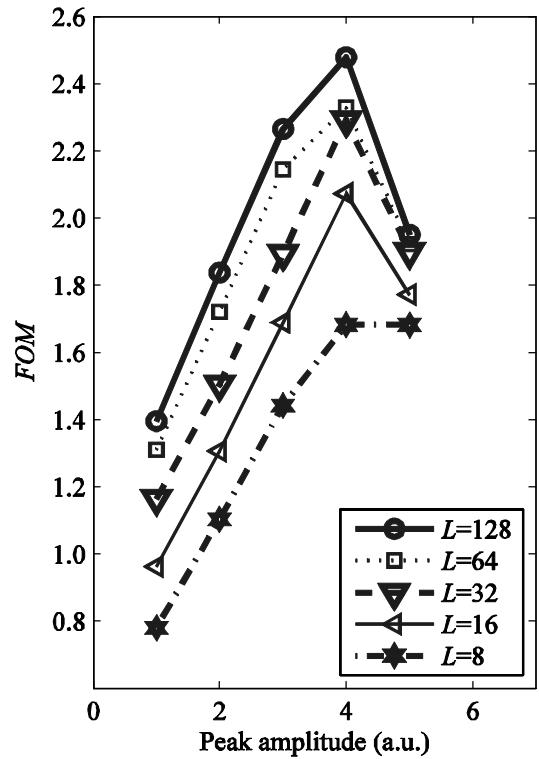

**Fig. 6** Plots of *FOM* versus peak amplitude for the five $L$ values.

To illustrate quantitatively how *FOM* depends on peak amplitude and $L$, Fig. 6 with plots of *FOM* versus peak amplitude for the five $L$ values is provided using the data shown in Fig. 4. The n/γ separation for the [4 6] bin of the peak amplitude especially declined for each $L$, which can be attributed to several nonlinear responses and/or non-stationary noise sources found in the scintillation detection system at higher peak amplitudes, i.e. higher incident particle energies.

## 5. Conclusions

The effect of baseline drift on the performance of the FGA method to discriminate neutrons and γ rays has been investigated in detail both theoretically and experimentally. Firstly, a data pre-processing method suitable for the FGA algorithm is proposed and described schematically by using a typical event. Secondly, the relationship between the baseline shift and the discrimination parameter is mathematically derived, from which we know that the more accurately we estimate the baseline shift, the smaller the error between $k_l$ and $k_t$ will be, and the better the discrimination performance of neutrons and gamma rays will become. Finally, the experimental data are processed and the result is consistent with that of the theoretical derivation.

After investigating the effect of the baseline shift on FGA, we decreased the complexity of the BLR approaches and thus reduced the cost on calculation and increased the counting rate of the measurement. This condition enhanced the applicability of FGA in the current embedded electronics systems to provide real-time discrimination in standalone instruments. We are currently studying the existing BLR approaches and applying the most appropriate one in FGA to substitute the programmable length mean filter used in this paper. Future work will also involve the exploration of more effective algorithms to estimate the baseline shift and validate the algorithms with more measurement data.

## 6. Acknowledgments

We acknowledge the support of the Institute of Nuclear Physics and Chemistry, the Chinese Academy of Engineering Physics, Mianyang, China. We also gratefully acknowledge the helpful discussions and advice from Dr. Li An, Dr. Pu Zheng and their technical team at the Chinese Academy of Engineering Physics, Mianyang.